\documentclass[letterpaper]{mn2e}
\title{The CLASS BL Lac sample: The Radio Luminosity Function}
\author[M. J. M. March\~a \& A. Caccianiga] 
       {M. J. M. March\~a$^{1,2}$ \thanks{e-mail:maria.marcha@gmail.com} \& A. 
Caccianiga$^{3}$ \thanks{e-mail:alessandro.caccianiga@brera.inaf.it} \\ 
$^1$ University of Lisbon, Portugal \\ 
$^2$ current address: 32 Ordnance Hill, London NW8 6PU, U.K. \\ 
$^3$ INAF - Osservatorio Astronomico di Brera, via Brera, 28, 20121 Milano, Italy \\ 
}
\begin{document}

\input{psfig.tex}

\date{Accepted 2013 January 09}
\pagerange{\pageref{firstpage}--\pageref{lastpage}}
\pubyear{2012}

\maketitle

\label{firstpage}

\begin{abstract}

This paper presents a new sample of BL Lac objects selected from a
deep (30 mJy) radio survey of flat spectrum radio sources (the CLASS
blazar survey, henceforth CBS). The sample is one of the largest well
defined samples in the low power regime with a total of 130 sources of
which 55 satisfy the 'classical' optical BL Lac selection criteria,
and the rest have indistinguishable radio properties. The primary goal
of this study is to establish the Radio Luminosity Function (RLF) on
firm statistical ground at low radio luminosities where previous
samples have not been able to investigate. The gain of taking a peek
at lower powers is the possibility to search for the flattening of the
LF which is a feature predicted by the beaming model but which has
remained elusive to observational confirmation. In this study we
extend for the first time the BL Lac RLF down to very low radio powers
$\sim 10^{22}$W/Hz, ie, two orders of magnitude below the RLF
currently available in the literature. In the process we confirm the
importance of adopting a broader, and more physically meaningful set
of classification criteria to avoid the systematic missing of low
luminosity BL Lacs.  Thanks to the good statistics we confirm the
existence of weak but significant positive cosmological evolution for
the BL Lac population, and we detect, for the first time the
flattening of the RLF at $L \sim 10^{25}$W/Hz in agreement with the
predictions of the beaming model.

\end{abstract}


\section{Introduction}

BL Lacs are flat radio spectrum AGN which show high variability in
both flux and polarization and whose Spectral Energy Distributions
(SED) are dominated by non-thermal processes. In these features BL
Lacs are not alone. In fact, as early as the first conference dedicated
to them as a class (Wolfe, 1978), the new term 'blazar' emerged, thus
grouping BL Lacs and Flat Spectrum Radio Quasar (FSRQs). 
Though this early amalgamation aimed at a
unified model that would explain both types of sources, BL Lacs have
nevertheless remained an elusive type of AGN throughout. For example,
due to the lack of strong emission features in their optical spectrum,
several BL Lacs are still without a redshift measurement, despite
continued efforts to obtain them (see for instance, Meisner \& Romani,
2010; Rau et al., 2012; Sbarufatti et al., 2009). Another aspect of
the enigmatic nature of BL Lacs is the fact that there have been
claims that a fraction of these sources suffers negative cosmological
evolution (see for instance Morris et al., 1991; Wolter et al., 1994;
Bade et al, 1998; Giommi, Menna \& Padovani, 1999; Rector et al.,
2000; Giommi et al., 2012) something that is in marked contrast with
the other blazars.

Though there is good observational evidence to support the generally
accepted view that the 'blazar phenomenon' is the consequence of
observing radio galaxies down their relativistic jets (see for
instance Urry \& Padovani 1995 and references therein; Capetti et
al. 2002; Trussoni et al. 2003; Giommi et al., 2012, there are still
detailed issues to be resolved concerning this 'zeroth order
unification'. In particular it is important to keep in mind that
though progress has been made in the analysis and high frequency
observations of this type of AGN, it is still true that there is a
lack of well defined samples that allow the estimate of statistically
important parameters such as the luminosity function (LF) and
cosmological evolution, especially in the low luminosity regime. For
instance, even though there are a few thousand of BL Lacs known
(Massaro et al., 2011 mentions roughly 1200 with the last version of
the BZ catalogue), the number of sources belonging to well defined
samples from which statistical results can be obtained is much smaller
than this.  Furthermore, until recently these samples did not sample
the population homogeneously in their radio luminosity. In fact, the
high radio luminosity end of the population was almost exclusively
sampled via radio surveys, specifically the 1Jy sample (Stickel et
al., 1991), whereas the lower radio luminosity samples were either
X-ray selected or obtained through a combination of X-ray and radio
data (see for instance Laurent-Muehleisen et al., 1998 and 1999 for the RGB;
Perlman et al., 1998 for the DXRBS; Caccianiga et al., 2002b for
REX). Since both types of surveys are sensitive to different parts of
the SED, studying the entirety of blazar phenomenon could be
compromised by selection biases due to the different ways the objects
in low and high luminosity regime were selected (eg. see the
discussion in Caccianiga \& March\~a, 2004).

Clearly the need to have larger samples that provide blazar data
across the radio luminosity range is acknowledged (see for instance
Giommi et al., 2012 for a discussion). The problem however resides in
the very nature of BL Lacs since the lack of strong emission features
in their optical spectra means that BL Lac samples are plagued with
selection effects. There are two main reasons for this: (1) the
initial BL Lac optical criteria required the EW of the strongest
emission lines to be below 5\AA, something that is not only arbitrary,
but also without much physical meaning, and (2) the fact that how weak
a line is to be recognized before being dismissed depends on a
combination of instrumental and intrinsic source parameters. Several
studies investigated the importance of the biases introduced in BL Lac
samples due to these type of selection effects, and updated selection
criteria considered more physically meaningful were introduced (Browne
\& March\~a, 1993, March\~a \& Browne, 1995, March\~a et al., 1996,
Landt et al., 2002, 2004).

Making use of the expanded selection criteria new BL Lac samples have
been produced (see for instance Giommi et al. 2012, for a discussion
of different samples). The advantage of these samples has been to
broaden the parameter space within which the BL Lac population could
be studied. Nevertheless, important parameters such as the Radio
Luminosity Function (RLF) of BL Lacs has only been possible in two
previous samples, both of which contain a small number of sources - 34
sources in the case of the 1Jy sample of Stickel et al., 1991, and 24
objects in the case of the DXRBS from Padovani et al., 2007 - and
reaching radio luminosities of around $10^{25}$W/Hz in the first
instance, and $10^{24}$W/Hz in the second.

In the current work we present a large and well-defined radio selected
sample of low power BL Lacs. We use it to investigate the cosmological
properties of this class of AGN, and to derive the Radio Luminosity
Function (RLF) to radio luminosities as low as $10
^{22}$WHz$^{-1}$. The paper is organised as follows: The next Section
discusses the selection of the CLASS sample of blazars and how we used
the available observational parameters to improve the BL Lac
classification at low radio luminosities. Section 3 is dedicated to
assessing cosmological evolution, whereas Section 4 sees the analysis
of the RLF for the samples identified in the previous
sections. Finally, in Section 5, we discuss the results and
conclusions are drawn. Throughout the paper we have assumed $S_{\nu}
\propto \nu^{-\alpha_{\nu}}$, with $\alpha_{o}=1.0$ for the optical
spectral index, and $H_{0}=71$ Km s$^{-1}$Mpc$^{-1}$. However, when it
came to comparing our results to those of previous studies we adopted
$H_{0}=50$ Km s$^{-1}$Mpc$^{-1}$, as some of the reference quantities
have been determined with this value. We have also adopted an
'old fashioned' cosmology $\Omega_{m}=0$, $\Omega_{\lambda}=0$ as
these will have little effect on the low redshift regime we are
primarily investigating.

\section{The CLASS sample of blazars}

The CLASS blazar sample (CBS) consists of 302 flat spectrum radio
sources selected according to the following criteria:

\begin{enumerate}
\item {$35\degr \leq\delta\leq75\degr$}
\item {$|b^{II}|\geq$20$^\circ$}
\item {$S_{5}\geq$ 30 mJy}
\item {flat spectrum, i.e. $\alpha_{1.4}^{4.8}\leq$0.5
(S$_{\nu}\propto\nu^{-\alpha}$)}
\item {red magnitude $\leq$ 17.5}\footnote{Such a limit on the  magnitude of the sources ensures that we target low redshift and low
  radio power sources}
\end{enumerate}

The details of the selection, as well as the radio properties of the
sample are discussed in March\~{a} et al. (2001) and the reader is
referred to that work for more detailed information. Further work on
radio imaging and multi-frequency data of the sample enabled a
refinement of the blazar classification to 244 (decreased from 302) of 
the objects in the CBS. This analysis is discussed in Caccianiga \&
March\~a (2004).

Optical classification has been collected, either from the literature
or from specific observations for about 91\% of the objects in the
sample. A discussion on the spectroscopic classification is presented
in Caccianiga et al. (2002a) but essentially,
objects have been broadly divided into three categories: Type~0 if
their spectra show only weak or no emission lines (see next
Section); type~1 if the spectra show broad emission features, and
finally, type~2 if the spectra only show narrow emission lines. 

The focus of the current work is on BL Lacs, hence on sources whose
spectrum is devoid of strong emission lines. However, at lower radio
luminosities isolating these sources is not as trivial as it seems
when dealing with more powerful objects. In particular, two problems
make this task more difficult: (1) the lack of an absolute measure of
'weak' and 'strong', and (2) the difficulty of recognising weak BL Lac
nuclei in the core of luminous host galaxies. In the absence of
physically meaningful cut-off criteria different authors have made
different attempts to minimise the effects of these two problems on
the selection of a given sample. The approach followed in the present
work is to investigate different subsamples resulting from applying
variations on the selection criteria that have been used in the case
of previous samples of BL Lacs, and investigate how that affects the
results and interpretation of their cosmological properties. A Table
(\ref{tab.cbsbllac}) containing the relevant parameters of the sample used in
the present work can be found in appendix A.

\subsection{The 'classical' BL Lacs}

Traditionally, a source is classified as a BL Lac if its optical
spectrum is featureless or shows only weak emission lines. A working
upper limit on the Equivalent Width (EW) of the strongest emission
line has been set to 5\AA\ in order to separate BL Lacs from other
radio-loud AGN. An additional criterion for the classification of BL
Lacs in the EMSS survey was introduced by Stocke et al. (1991).  It
consisted in requiring that enough non-thermal continuum from the AGN
was present in addition to the starlight of the host galaxy so that
the observed contrast across the CaII H\&K discontinuity at 4000\AA
($\Delta$) was reduced to $\leq$ 25\%.  However, the imposition of
such a limit on the relative strength of the optical flux of the AGN
to the starlight of the host galaxy is somewhat arbitrary and unlikely
to be related to the intrinsic properties of the active nucleus. In
fact, the measured contrast is inversely proportional to the strength
of the non-thermal contribution from the AGN, which means that as the
AGN gets weaker, the measured contrast will get larger. Hence, the
deeper the survey, the more important is the choice of contrast used
in the selection of BL Lacs. A too limiting value for this quantity
can introduce significant incompleteness in the final sample as is
widely discussed in Browne \& March\~a (1993) and March\~a \& Browne
(1995).

It was first proposed by March\~a et al. (1996) while studying a
sample of flat radio spectrum sources with flux limit of 200mJy that
the limiting contrast for BL Lac definition should be extended to 40\%
in order to avoid missing weak nuclei in the core of bright
ellipticals. This was supported by the lack of strong evidence to
separate sources below and above the 25\% threshold in a sample of
optically bright, low luminosity flat radio spectrum objects. Since
then, various works have lent further support for this expanded
classification criterion, and a limiting contrast of 40\% has now been
widely used in the definition of new BL Lac samples
(Laurent-Muehleisen et al. 1998; Caccianiga et al.  1999, Perlman et
al., 1996: Beckmann et al., 2003; Landt et al. 2002, Padovani et al.,
2007).

\subsection{The Type~0 sources}

Even though the inclusion of sources with $\Delta \leq 40\%$ in the BL
Lac classification prevents significant incompleteness of radio
samples down to a flux limit of $\sim$200mJy, it is not guaranteed 
that this limit will prevent weak BL Lac nuclei from being detected
in much deeper surveys like the CBS where most of the selected low
power BL Lacs are likely to be indistinguishable from normal
``passive'' elliptical galaxies. For this reason, we chose to study
also all the sources in the CBS that are spectroscopically
classified as Type~0, i.e, sources showing an optical spectrum with
weak (EW$\leq$5\AA) or no emission lines, and no restriction on $\Delta$

There are 93 sources that fall into this category: the 55 'classical'
BL Lacs plus 38 sources named PEGs (Passive Elliptical Galaxies) which
apart from measured contrasts above 40\%, are otherwise
indistinguishable from BL Lacs. There are good reasons for doing this
as previous data such as the analysis of SEDs (Ant\'on et al. 2004;
Ant\'on \& Browne, 2005), radio imaging and polarization studies
(Dennett-Thorpe \& March\~a, 2000; Bondi et al., 2004), where at least
in the case of some PEGs, it was not possible to distinguish the
morphology and polarization properties from those of `classical' BL
Lacs. This may indeed be the best observational evidence yet to
support the existence of a BL Lac nucleus in the core of at least some
PEGs. In the case of the CBS it is more difficult to obtains such a
direct confirmation of the presence of a BL Lac nucleus in PEGs due to
the weakness of the selected sources. However, we have presented
evidence, based on high resolution VLA maps, that this is likely to be
true also in this sample (see Caccianiga \& March\~a, 2004).

\subsection{The Weak Emission Line sample}

The optical criteria separating BL Lac objects from quasars is known
to be somewhat arbitrary, and in the two previous sections we
discussed ways to prevent incompleteness in BL Lac samples due to this
arbitrariness especially when one has the advantage of starting from a
radio survey. Other attempts to address the optical identification of
BL Lacs have however been published in the literature, and it is
important to investigate how they may compare to those discussed in
the two previous Sections.

In 2004, Landt et al.  adopted a diagnostic plot based on the
equivalent width of two narrow emission lines, namely the
[OII]$\lambda$3727\AA\ and the [OIII]$\lambda$5007\AA. According to
these authors, this diagnostic plot is able to separate AGN that are
intrinsically ``weak lined'' (WLAGN) from the ``strong lined'' ones
(SLAGN) independently to their orientation.

For consistency with what is presented in the literature we want to
apply also this alternative classification scheme to the CBS sample to
produce the RLF of the population of WLAGN which should be considered
an extension of the population of type~0 based on a more ``physical''
classification.

\begin{figure}
\psfig{figure=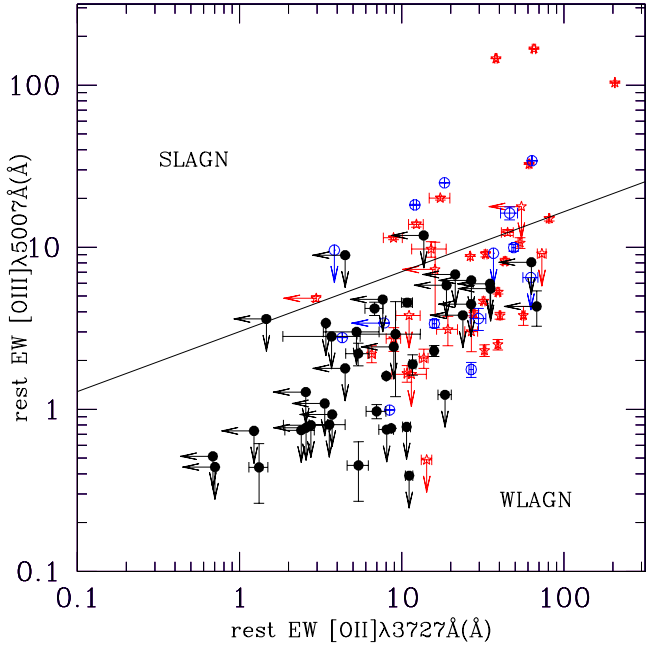,width=9cm}
\caption{The classification plot proposed by Landt et al. (2004) applied
to the CBS blazars with large values of 4000\AA\ break ($\Delta>$20\%). 
The solid line is the dividing line proposed by Landt et al. (2004)
to separate between WLAGN and SLAGN. 
Symbols indicate the former classification discussed in Caccianiga et al. 
(2002a): filled circles = type~0, open circles = type~1, stars = type~2}
\label{landt}
\end{figure}

Only for $\sim$ half of the 244 CBS sources classified as "blazar",
based on radio data (see Caccianiga \& March\~a 2004 for details), do
we have a spectrum covering both the [OII]$\lambda$3727\AA\ and
[OIII]$\lambda$5007\AA\ emission lines used in the work of Landt \&
co-workers. The sources for which we have not covered one of the two
lines (or both) are typically high z objects (z$>$0.8).
However, the choice of the classification scheme is really critical
for blazars with a relatively low luminosity (and, hence, low
redshift), where the possible effects due to the orientation may be
relevant. For powerful blazars the Landt et al. (2004) converges to
the usual classification based on the EW=5\AA\ threshold. An indicator
of the AGN optical luminosity is the value of the 4000\AA\ break
($\Delta$): sources with $\Delta<$20\% are usually powerful blazars
(typically with P$_{5 GHz}>$10$^{25}$ W Hz$^{-1}$) for which the Landt
et al. (2004) classification in WLAGN/SLAGN is almost equivalent to
the Type~0/Type~1 classification that we have adopted in the CBS
sample. For these sources with $\Delta <$20\%, we have thus simply
``converted'' our classification into the WLAGN/SLAGN classification
by assuming type~1=SLAGN and type~0=WLAGN. For the sources with
$\Delta>$20\%, instead, we have applied the
[OII]$\lambda$3727\AA/[OIII]$\lambda$5007\AA\ classification plot
proposed by Landt et al. (2004) to distinguish between WLAGN and
SLAGN.  Since these sources are typically at low redshifts (z$<$0.3)
both emission lines are covered by our spectra in the large majority
of cases ($\sim$80\%). For the remaining objects the missing emission
line is typically the [OII]$\lambda$3727\AA\ because the available
spectrum does not cover the blue part.  As discussed in Land et
al. (2004) the [OIII]$\lambda$5007\AA\ alone can be used in these
cases to provide a tentative classification: if it is large
($>$20\AA), then the source is likely a SLAGN while, if it is small
($<$3\AA), the source is likely a WLAGN. In the intermediate cases the
ambiguity can be removed only by measuring the [OII]$\lambda$3727\AA\
equivalent width, so we consider these sources as
``unclassified''. Only 17 blazars, out of 244 (7\%), belong to this
group of unclassified sources.  In Fig.~\ref{landt} we report the
classification plot for the CBS sources with $\Delta>$20\% and for
which both emission lines are covered. Most of these objects fall in
the WLAGN region.

In total, the sample of WLAGN includes 129 objects. This sample
contains all but one of the type~0 objects, described in the previous
section, plus some low-power sources previously classified as type~1
or type~2 objects. We have also decided to include the two Type~0 objects
whose upper limits on the EW of the two lines lie above the dividing
line in Figure ~\ref{landt}) as WLAGN.

\section{Assessing cosmological evolution}

The evolutionary properties in this work are investigated via the
`$V/V_{max}$ test'. First introduced by M. Schmidt (1968) to study the
space distribution of quasars the test was later generalised by Avni
\& Bachall (1980) for the case where the sample selection involved
more than one survey.  Basically, the test consists in assessing the
ratio between the volume $V$, corresponding to the volume ''enclosed''
by the redshift of the source, and the maximum volume ($V_{max}$)
within which the source can still be found satisfying the selection
criteria of the survey. In the specific case of the CBS there are two
flux limits that need to be considered, one in the radio and one in
the optical, and consequently, the $V_{max}$ to be used will be the
one corresponding to the most limiting of the two.

For non-evolving populations, the ratios $V/V_{max}$ computed for each
source should be uniformly distributed between 0 and 1 with a mean
value $<V/V_{max}>=0.5$ and statistical errors being given by $\sigma
= 1/\sqrt(12N)$ (where $N$ is the number of sources in the sample).
If, on the other hand, the distribution of sources is non-uniform,
$<V/V_{max}>\neq 0.5$, and the population is said to be evolving:
positively if the mean is larger than 0.5, or negatively if it is
smaller than 0.5. It is however, important to emphasise the fact that
a deviation from a uniform distribution of $V/V_{max}$ can also
originate, not due to evolution but rather from unrecognised selection
effects in the sample. To distinguish between the two situations is
particularly relevant in the case of BL Lacs, where primarily due to
the properties of their optical spectra, weak nuclei  can easily
be missed in flux limited samples.

Before proceeding with the analysis, there is one issue that must be
addressed: how to deal with the $z$-less BL Lacs. There are 16 such
sources which is a significant fraction of the 55 BL Lac sample. We
have explored a number of possibilities for attributing a redshift to
each of the 16 $z$-less BL Lacs. We started out by considering just
the subgroup of 39 BL Lacs with $z$ measurements and then the
following possibilities: (1) all $z$-less sources were assigned a
$z=0.3$ (the mean of the BL Lacs with measured redshift, (ii) a
$z=0.8$ for all the $z$-less BL Lacs, and finally (iii) a random $z$
was found for the 16 $z$-less sources
\footnote{This was done by allowing the sources to have radio
  luminosities in the range typical of that observed for other radio
  selected BL Lacs, e.g. L1: $10^{24} \leq L_{r} \leq 3 \times 10
  ^{28}$~W/Hz with an extra condition of $z > 0.1$ since below this
  value (assuming a typical radio-to-optical ratio for BL Lacs), we
  should detect the host galaxy in the optical image. We have taken
  the exercise a step further and actually investigated two further
  luminosity intervals from where the random $z$ distribution of the
  $z$-less objects were drawn: L0:$10^{23} \leq L_{r} \leq 3 \times 10
  ^{28}$, and L2 :$10^{25} \leq L_{r} \leq 3 \times 10 ^{28}$~W/Hz. We
  found that apart from the highest $z$ achieved ($z_{max} \sim 3$ for
  L0, to $z_{max} \sim 1.5$ for L2), it had little influence on the
  $<V/V_{max}>$ test, e.g. mean and distribution. We have used the
  interval L1 as a default but will discuss the implications of using
  the other luminosity intervals when analysing the LF.}  The results
  can be found on Table \ref{tab.zvar}. Essentially, there is minimal
  impact on the $<V/V_{max}>$ as a result of assigning different
  values of $z$ to the $z$-less sources. Consequently, for the
  remainder of this work we have considered the option where the 16
  $z$-less BL Lacs have been assigned a random value for their
  redshift (drawn from the L1 interval as defined in the footnote). We
  believe that using a random distribution for the $z$-less sources
  is, not only a more realistic approach than using the mean value of
  the sample for all the objects without redshift and one that is
  likely to have the least fictitious effects on the estimated RLF,
  but also one that is in agreement with a recent work by Rau et
  al. (2012) which suggest that some $z$-less BL Lacs have photometric
  redshifts as high as $z \sim 2.$.

\begin{table*}
\begin{tabular}{ccccc} 
  & BL Lacs w/ $z$ & $z$-less$=0.3$ & $z$-less$=0.8$ & $z$-less$=$random \\ 
$<V/V_{max}>$ & 0.59 & 0.62 & 0.63 & 0.62  \\ 
& & & & \\
 $1\sqrt(12N)$ & $\pm$0.05 & $\pm$0.04 & $\pm$0.04 &  $\pm$0.04 \\
\end{tabular}
\caption{Mean $V/V_{max}$ values for the BL Lac CBS sample for
different redshift options for the $z$-less sources.}
\label{tab.zvar}
\end{table*}

We have analysed 4 subsamples separately, 3 from the CBS - BL Lacs,
Type0, WLAGN - plus the 1Jy BL Lac sample of Stickel et al.,
1991 as it provides a validity test on the analysis. 

\subsection{Do BL Lacs  evolve?}

The cosmological evolution of BL Lacs has been a matter of debate over
the last two decades and it is still considered as an open problem. At
the origin of the debate are the opposite results obtained when using
the only 2 statistically complete samples of BL Lacs available in the
early 90s, one selected at radio frequencies (1 Jy sample;
Stickel et al. 1991) and the other in the X-ray (the EMSS-C sample;
Morris et al. 1991). In the case of the radio selected sample the
result was consistent with a population that suffered no evolution, or
a mildly positive one, whereas in the case of the X-ray selected
sample the result was clearly indicative of a population that evolved
negatively.  The evolutionary picture of BL Lacs became even less
clear when later results based on ROSAT data (Bade et al. 1998;
Giommi, Menna \& Padovani 1999; Beckmann et al., 2003; Giommi et
al. 2001) seemed to confirm the presence of negative evolution for a
subset of the BL Lac population, i.e. for the High energy peaked BL
Lacs (HBL), while other sets of data, namely the DXRBS and the REX
surveys found no evidence for evolution (Caccianiga et al., 2002b;
Padovani et al., 2007).

More recently, Giommi et al., (2012), suggest that BL Lacs are
essentially of two types: those that are beamed cores of FRII radio
galaxies and which should be grouped with FSRQs, thus sharing their
strong cosmological evolution, and those that show no evolution
because they are beamed versions of FRIs. The authors performed
extensive numerical simulations and found that if they consider only
the most extreme HBLs, then an apparent negative cosmological
evolution appears as a result of a combination of different selection
effects. Can the CBS sample help shed any light onto this matter?

Table \ref{evol.tab} shows the values obtained for the $<V/V_{max}>$
test for each of the samples mentioned before. It also shows the
results of modelling the evolution with a simple exponential form
$\exp(C\times\tau(z))$ where $\tau(z)$ is the look-back time in units of
$H_{o}$. The values of the K-S test for the distribution of
$V/V_{max}$ before and after the evolution model are also
shown. 

The results can be summarised as follows:

\begin{enumerate}

\item We used the 1Jy sample of BL Lacs of Stickel et al. (1991) for a
  reliability check on our analysis and we find similar values for all
  the parameters (the same $<V/V_{max}> =0.6 \pm 0.05$, and $C=3$ vs. $C=3.1$
  cited in Stickel \& co-workers). More interestingly, the sample shows the
  weakest statistical need to invoke evolution. With a 2$\sigma$
  detection and a $V/V_{max}$ distribution that cannot be dismissed as
  incompatible with a uniform one, it is the sample where a LE model
  was fitted with the lowest exponential $C$ parameter.

\item The CBS samples of 'classical' BL Lacs, Type0 and WLAGN show similar
$<V/V_{max}>$ values and $V/V_{max}$ distributions although BL Lacs
showed the highest, and WLAGN the lowest deviation from the mean value
of 0.5. In both the CBS 'classical' BL Lac and Type0 samples the
deviation is now at 3$\sigma$ whereas for the WLAGN the deviation is
only detected at 2$\sigma$. Also, the $V/V_{max}$ distributions are now
inconsistent with being uniform, contrary to the case in the 1Jy sample.

\item Modelling the evolution yielded similar values for an LE
exponential model although the BL Lacs required the highest evolution
parameter and the WLAGN the lowest. We note however, that the
evolution parameter $C$ that achieves $<V/V_{max}>=0.5$ within 1$\sigma$
is pretty similar in the 3 cases;

\item How do these values compare with other works? The values of
$<V/V_{max}>$ obtained for the CBS samples of BL Lacs, WLAGN and Type0
are comparable to the one found for the 1Jy sample of BL Lacs but
whereas in the latter the deviation was only detected at 2$\sigma$, in
the case of the CBS samples the numbers allow a 2 or 3$\sigma$
confidence (depending on the sample in question) in the claim of
cosmological evolution. Furthermore, the $V/V_{max}$ distributions are
not consistent with a uniform one, in contrast to what happens in the
1Jy sample. We also found that the evolution parameter values required
to bring the $<V/V_{max}>$ down to 0.5 were larger in the CBS samples
($C \sim 5 - 6$) when compared to that obtained for the 1Jy sample ($C\sim 3$).

Apart from the 1Jy sample of Stickel et al. (1991), the closest
comparative study is the one carried out by Padovani et al., 2007 on
the DXRBS blazar sample. In this study Padovani and co-workers find
that the 24 BL Lacs that constitute the sample show $<V/V_{max}> =
0.57 \pm 0.06$, a value consistent with the one found for the CBS BL
Lacs. We therefore conclude that the BL Lac population suffers weak
(when compared to FSRQs) but significant positive evolution, a result
that was difficult to obtain due to the small statistics of previous
samples.

\end{enumerate}

\begin{table*}
\begin{tabular}{ccccccc} 
 Sample & $<V/V_{max}>$ & K-S (\%) & evol. model & $C$ &  $<V/V_{max}>$ & K-S (\%)  \\ 
\\

BL Lacs & 0.62 ($\pm0.04$) & $<<$1\% & LE(exp) & 6.$^{+1.}_{-1.5}$ & 0.50 & $\sim$60\%  \\
\\

Type~0  &  0.59 ($\pm0.03$) & $<<$1\% & LE(exp) & 5.5$^{+0.5}_{-1.5}$ & 0.50 & 70\%   \\

\\

WLAGN   &  0.57 ($\pm0.03$) & $<$1\%  & LE(exp) & 5.$^{+1.}_{-1.}$  & 0.50 & 99\%  \\

\\


\end{tabular}

\caption{Summary of the parameters for the evolutionary model in the
case of the different samples. Columns are: (1) Sample; (2)
$<V/V_{max}>$ obtained and its respective error; (3) K-S test
probability for the $<V/V_{max}>$ (see text for details); (4) type of
evolutionary model used, (5) Evolutionary parameter where the upper
and lower limits indicate the $C$ which still falls within
$<V/V_{max}>=0.5 \pm 1 \sigma$; (6) K-S test probability for the
$<V/V_{max}>$ distribution after the evolution was taken into
account.}

\label{evol.tab}
\end{table*}

\section{The Radio Luminosity Function}

The differential luminosity function is determined according to the
following equation:

\begin{equation}
\Phi_{L} = \frac{4\,\pi}{A}\, \frac{1}{\Delta L} \, \sum_{j=1}^{N} \,
(\frac{1} {V_{max}})_{j} \: ,
\label{lf}
\end{equation}

where $V_{max}$ is the maximum volume, A corresponds to surveyed area,
$\Delta L$ the width of the luminosity bin, and N is the number of
objects contained in the bin (L,L$+\Delta$L). As discussed previously
in Section 3, $V_{max}$ is the smallest maximum volume between those
computed using the radio and the optical limits of the sample. In this
way the presence of the optical magnitude cutoff is automatically
taken into account also in the derivation of the RLF.

\subsection{The analysis of the RLF}

The analysis of the LF holds vital statistical information about the
population being studied. For example, some authors claim that the
slope of the RLF of radio galaxies can give an idea of the rate at
which AGN producing jets appear in the universe (Kaiser \& Best,
2007). In the particular case of BL Lacs, the underlying assumption is
that these sources consist of beamed up cores of low luminosity radio
galaxies. In such a scenario the LF of the two populations should be
connected. Basically, if the LF of the parent population (unbeamed)
consists of a single power-law, the effect of beaming produces a break
in the LF at a luminosity that depends on the Doppler factor
$\delta=[\Gamma(1-\beta cos\theta)]^{-1}$, where
$\Gamma=(1-\beta^{2})^{-1/2}$ is the Lorentz factor, $\beta$ the jet
velocity in units of the velocity of light, and $\Theta$ the angle
between the jet and the line of sight (l.o.s). Detailed analysis of
the effect of beaming on the LF has been carried out by several
authors (see for example Urry \& Shafer, 1984 and Urry \& Padovani
1991 and 1995 for thorough discussions).

\begin{figure*}

\centerline{
\psfig{figure=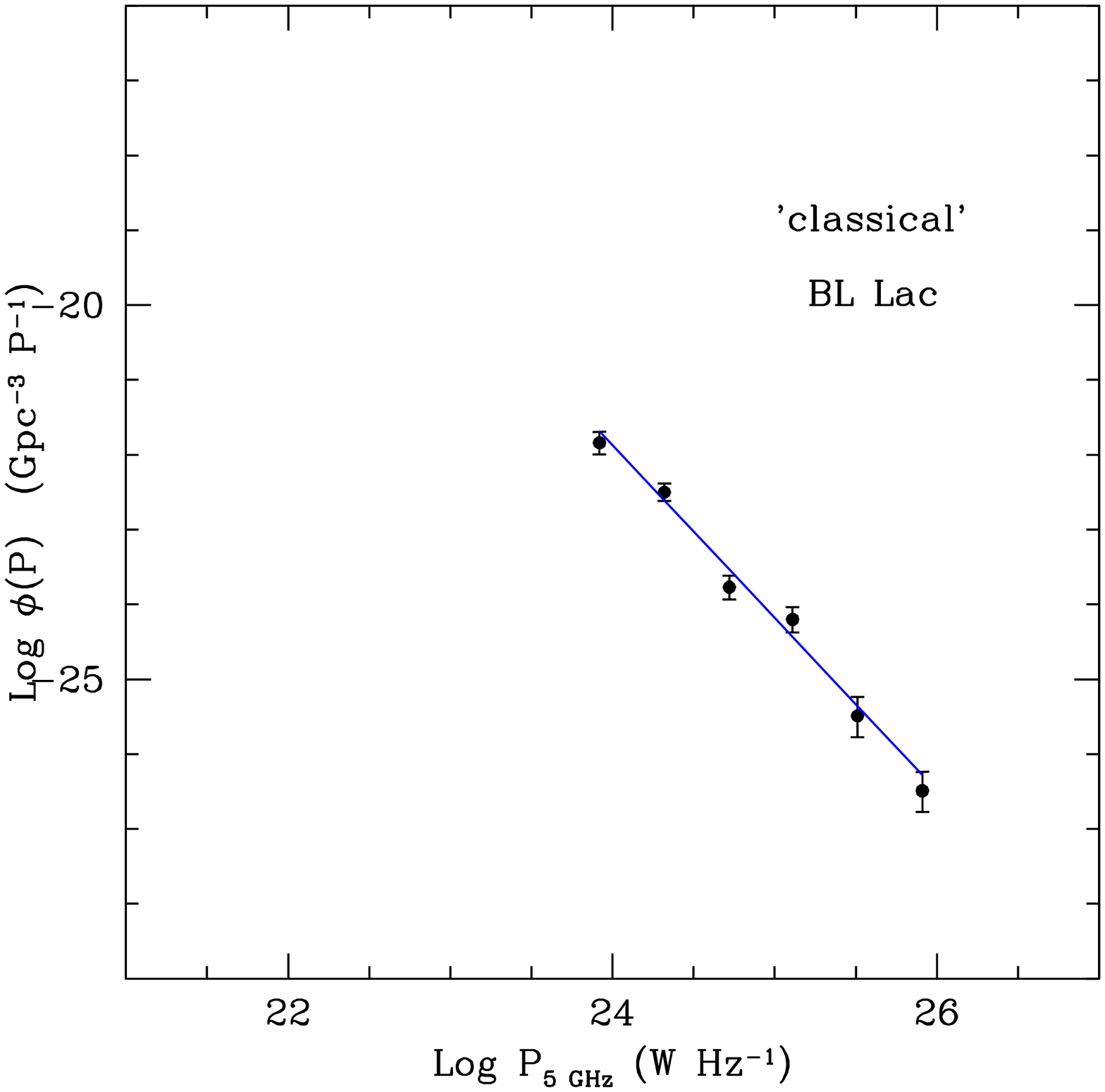,height=7cm,width=7cm}
\psfig{figure=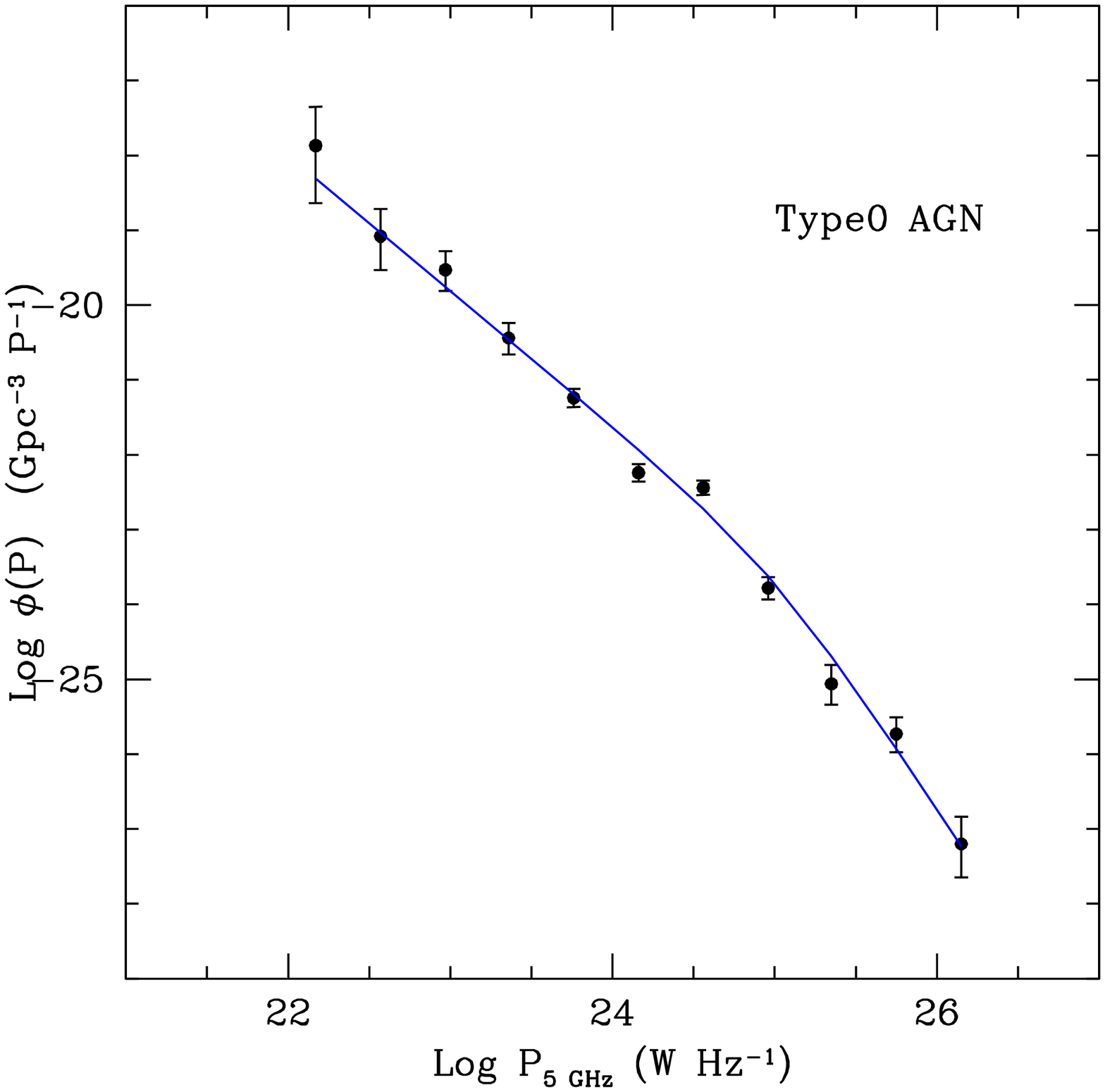,height=7cm,width=7cm}
}

\centerline{
\psfig{figure=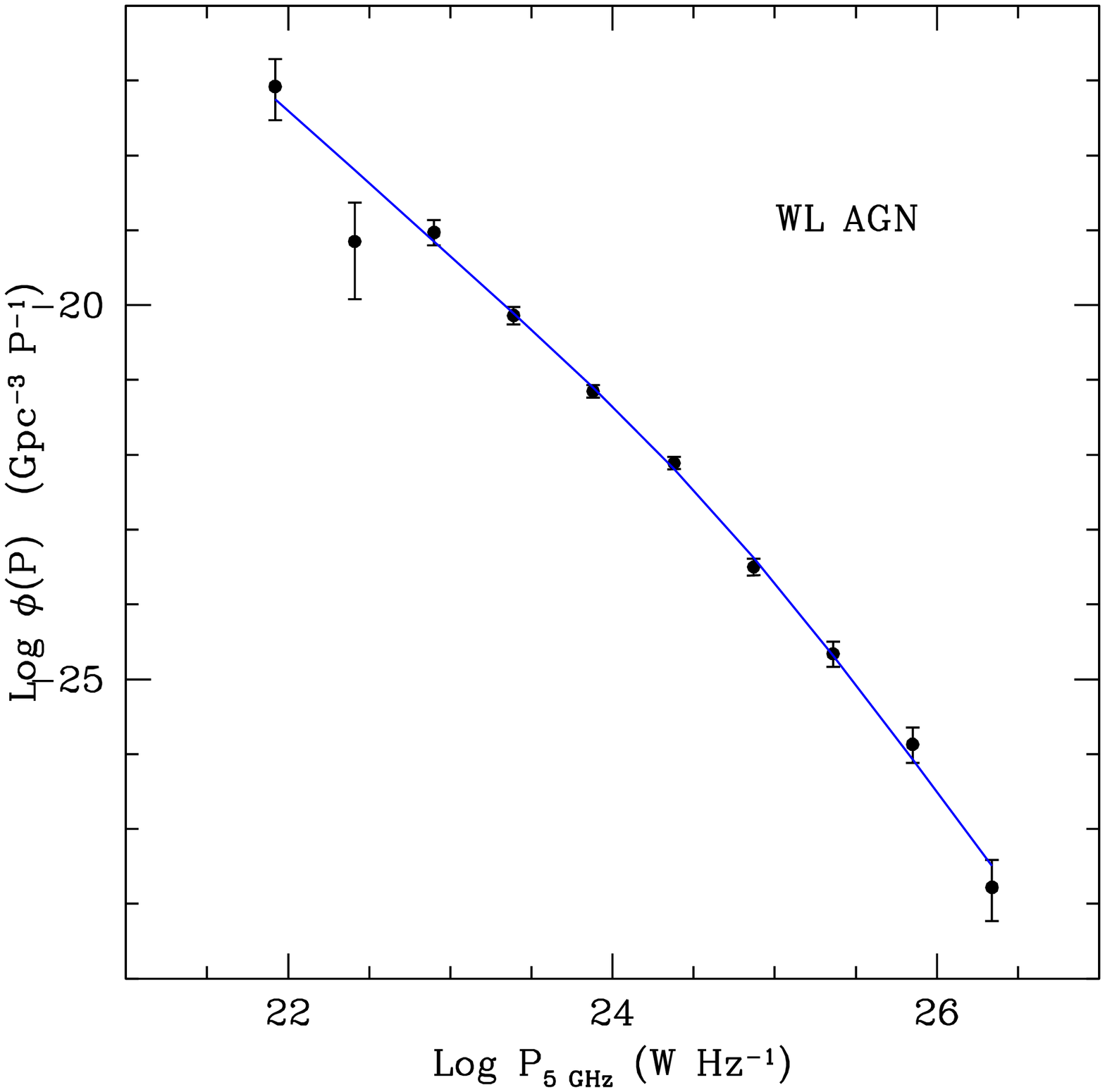,height=7cm,width=7cm}
}

\caption{Estimated RLF in the evolution scenario. From Top to bottom:
BL Lacs, Type0, WLAGN. The best fits are shown by the continuous
lines. ($H_{0}=71$ Km s$^{-1}$Mpc$^{-1}$)}

\label{lf.fig}
\end{figure*}

\begin{figure*}
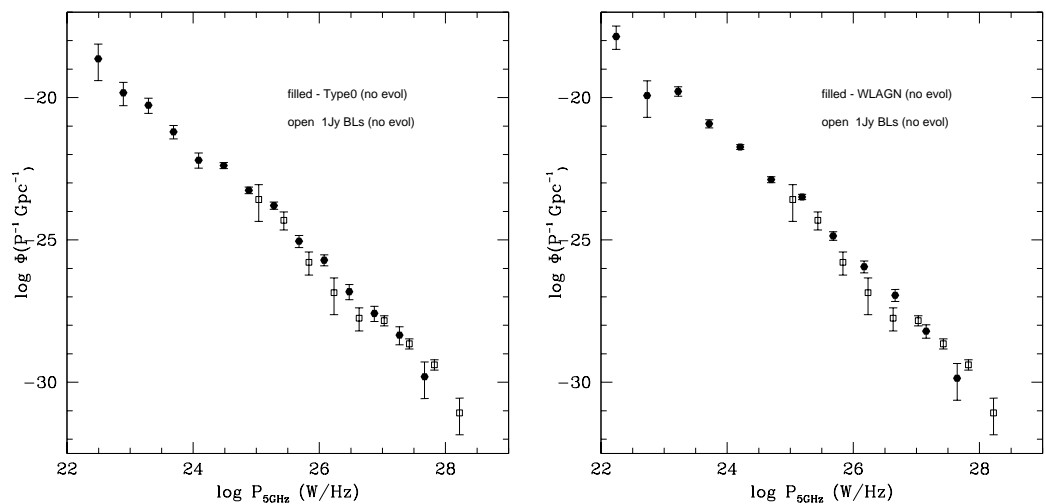


\centerline{
\psfig{figure=fig.lf.ty0w1jy.EV0,height=7cm,width=7cm}
\psfig{figure=fig.lf.wlagnw1jy.EV0,height=7cm,width=7cm}
}

\caption{RLF for the Type0 and WLAGN samples (filled circles) against
the 1~Jy sample (open squares) of BL Lacs in the no evolution
scenario. ($H_{0}=50$ Km s$^{-1}$Mpc$^{-1}$) }

\label{1jycomp.fig}
\end{figure*}

In the present work the RLF for the CBS will be studied under two
scenarios: evolution (with an exponential LE model), and
no-evolution. For each case, the LF was fitted by:

\begin{equation}
\Phi (L) = K  L^{a} 
\label{pl.eq}
\end{equation}

for a single power-law, or 

\begin{equation}
 \Phi (L) = \frac{ \Phi_{o}} { (\frac {L}{L_{break}})^{a} + (\frac {L}{L_{break}})^{b} }
\label{bpl.eq}
\end{equation}

for a broken slope at a given luminosity $L_{break}$. The results are
shown in Table \ref{lffit.tab}, and Figures \ref{lf.fig},
\ref{1jycomp.fig} and \ref{RLFcomp.fig}.

\begin{table*}
\begin{tabular}{cccccc} 
Sample & fit & (no evol.) & $\chi^{2}$/Dof & (evol) 
& $\chi^{2}$/Dof \\ 
\\ 

\hline 

BL Lacs & & & & C=6 & \\ 

p-law &  a   & -2.09$^{+0.08}_{-0.08}$ & 4/7 & -2.31$^{+0.12}_{-0.11}$ & 7/4\\ 
\\ 
broken p-law &  a & -- & -- & -- & -- \\ 
             & b & -- & & -- & \\ 
             & Log $L_{b}$(W/Hz) &  & &  & \\ 
             & Log $L_{b}$ range  &  & &  & \\ 
\\ \hline 

Type~0 & & & & C=5.5 & \\

  p-law   &  a  & -2.02$^{+0.05}_{-0.05}$  &  23/12 & -2.15$^{+0.07}_{-0.07}$  & 37/9  \\
\\
broken p-law & a & 2.49$^{+0.24}_{-0.18}$ & 10/10 & 3.31$^{+0.35}_{-0.32}$ & 21/7   \\
\\
  &       b & 1.66$^{+0.14}_{-0.17}$ &     &    1.81$^{+0.12}_{-0.13}$   &   \\
\\
  & Log $L_{b}$(W/Hz) & 25.09$^{+0.64}_{-0.67}$ & & 25.01$^{+0.25}_{-0.26}$  & \\
\\
  & Log $L_{b}$ range  & [24.75,25.21] &  &    [25.01, 25.67] &  \\ 

\\
\hline

WLAGN  &               &                &       &  C=5     &              \\    

  p-law       & a  & -2.07$^{0.04}_{-0.04}$ & 25/10 & -2.29$^{+0.05}_{-0.06}$  &  24/8 \\
\\
broken p-law & a  & 2.71$^{+0.47}_{-0.27}$ & 13/8 & 2.93$^{+0.33}_{-0.24}$  & 8/6   \\
\\
        &       b  & 1.87$^{+0.10}_{-0.12}$ &    & 1.93$^{+0.15}_{-0.19}$   &     \\
\\
  & Log $L_{b}$(W/Hz) & 25.69$^{+0.63}_{-0.76}$ & &  24.64$^{+0.51}_{-0.56}$ &  \\
\\ 
 & Log $L_{b}$ range  & [25.69,26.54] &    &    [24.64, 24.99] & \\ 
\\

\end{tabular}

\caption{Table containing the parameters for the fitting of the
LF. Columns are as follows: (1 and 2) - Sample and parameters for the
type of fit, either power-law or broken power-law; (3 and 5) - slope
of the LF, luminosity break, and the range of the luminosity break
obtained by varying the luminosity interval from which the random $z$
distribution was obtained, e.g L0-L2 as defined in Section 3.; (4 and
6) - $\chi^{2}$/degrees of freedom. }

\label{lffit.tab}

\end{table*}

The results can be summarised as follows:

\begin{figure*}

\centerline{
\psfig{figure=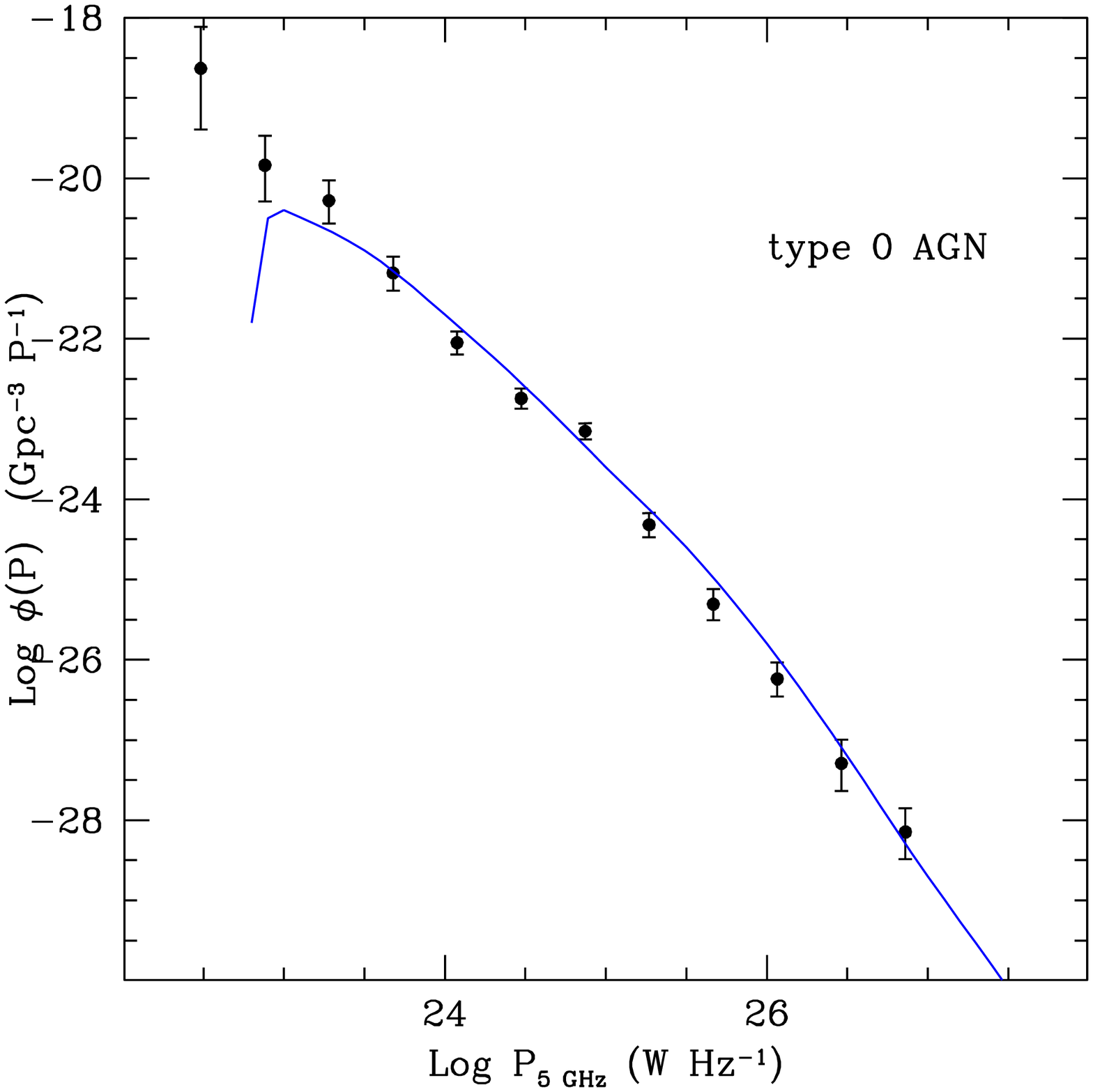,height=7cm,width=7cm}
\psfig{figure=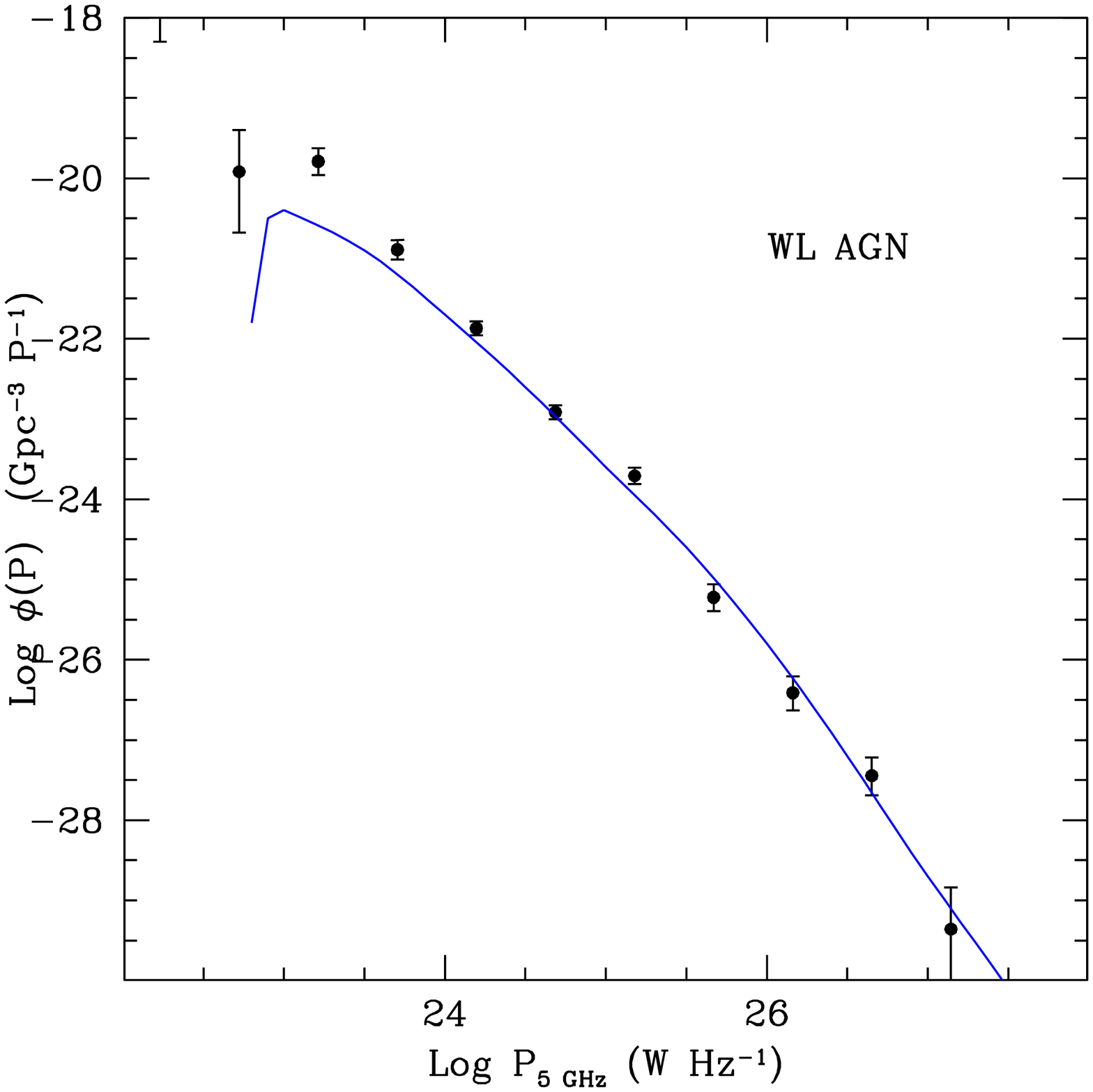,height=7cm,width=7cm}
}

\centerline{
\psfig{figure=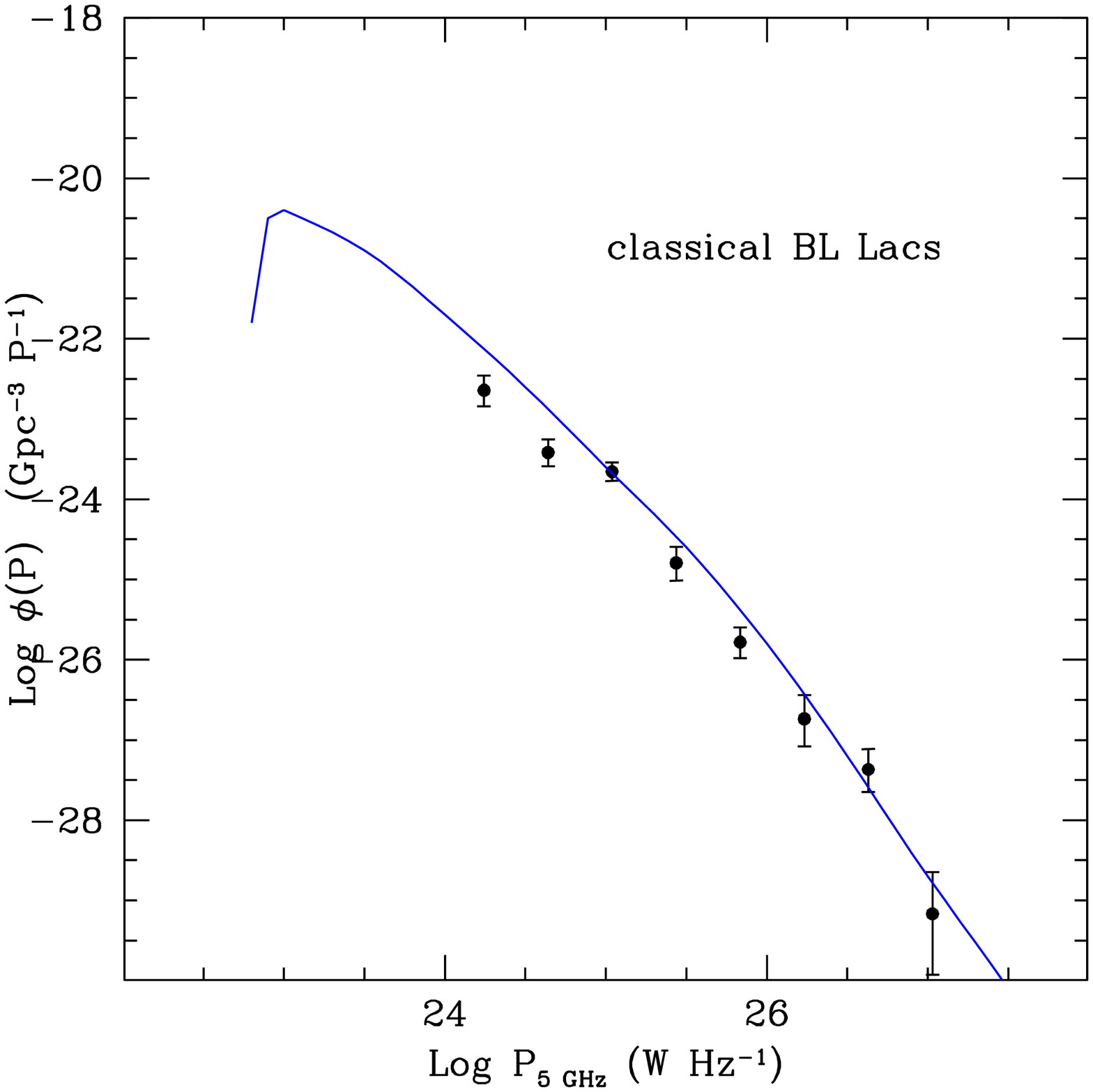,height=7cm,width=7cm}
}

\caption{ Points indicate the observed LF de-evolved according to an
exponential LE model with evolution parameter $C=3$ against the
predicted LF according to Urry, Padovani \& Stickel (1991) for the
Type0, WLAGN and 'classical' BL Lac samples discussed in this
work. ($H_{0}=50$ Km s$^{-1}$Mpc$^{-1}$) }

\label{RLFcomp.fig}
\end{figure*}

\begin{enumerate}

\item For the 'classical' BL Lacs of the CBS we find that the
  reasonable fits to the shape of the RLF are those of a single power
  law (see Figure \ref{RLFcomp.fig} and Table \ref{lffit.tab}). The
  slope in the evolution scenario is consistent with the steepest
  value found for the 1Jy sample of BL Lacs (Stickel et al., 1991
  found slopes between 2.4 and 2.7 depending on the binning), and
  those found for the DXRBS (slope of $\sim$2.3 for a bin size of
  $\Delta log P=0.6$). The evolution parameters are quite different in
  the three samples ($C=6$ in the case of the CBS and $C=3$ and $C\sim
  2.2$ for the 1 Jy and DXRBS samples, respectively). It is important
  to note that just as in the case of the DXRBS, the CBS 'classical'
  BL Lac sample achieves radio luminosities at least a factor of 10
  below those of the 1Jy BL Lac sample. The total number density found
  for the CBS 'classical' BL Lac sample from the integral LF for the
  evolution scenario is between those values found for the 1Jy and the
  DXBRS. Specifically, for the same cosmology ($\Omega_{m}=0$,
  $\Omega_{\lambda}=0$, and $H_{0}=50$ Km s$^{-1}$Mpc$^{-1}$) Stickel
  et al. (1991) found a value of $N \sim 40$ Gpc$^{-3}$ for the
  interval of luminosity interval $6 \times 10^{24}$~W/Hz - $3 \times
  10^{27}$~W/Hz, while Padovani \& co-workers found the total number
  density of the DXRBS BL Lacs to be $N \sim 310$ Gpc$^{-3}$ for the
  the luminosity interval of $10^{24}$W/Hz to 6$\times
  10^{26}$W/Hz. For the CBS 'classical' BL Lac sample this number was
  found to be $N \sim 90$ Gpc$^{-3}$ for luminosity range $10^{24}$W/Hz
  to 6$\times 10^{26}$W/Hz.

\item For Type~0 and WLAGN we find that the RLF is best fit by a
broken power-law (see Figure \ref{RLFcomp.fig} and Table
\ref{lffit.tab}). It is important to emphasize that the RLF for the
Type0 and the WLAGN show good overall agreement in the region of
overlap with the 1Jy sample (i.e., luminosities roughly between
$10^{25}$ and $3 \times 10^{27}$~W/Hz - see Figure \ref{1jycomp.fig})
but that their luminosity range extend to a factor of $\sim 1000$
times below. We note that such an agreement warrants credibility to
the methodology used to derive the statistical properties of the
sample. This is particularly relevant for the high luminosity end of
the LF which is the region where the CBS starts loosing coverage, and
one may worry about the sample not being representative, either
because of the magnitude limit in the selection criteria, or because
of the assumptions made for the sources without redshift
measurements. The fact that the CBS samples reach all but the very
last luminosity bin of the 1~Jy BL Lac luminosity function without any
real discrepancy between the two, gives good guarantee that any missed
sources at the high end of the luminosity range would have little
impact on the statistical analysis of the samples. The total number
density (for the evolution scenario and $H_{0}=50$ Km
s$^{-1}$Mpc$^{-1}$ cosmology considered before) derived from the
integral LF for the luminosity range $3\times10^{22}$ -
$3\times10^{26}$ W/Hz was found to be $N \sim 1.4 \times
10^{4}$Gpc$^{-3}$ and $N \sim 5 \times 10^{4}$Gp$^{-3}$ for the Type0
and WLAGN, respectively.

\item Taking the Type0/WLAGN we find for the first time evidence of
  the break in the LF according to the predictions of the Unified
  Schemes. Indeed, the broken power law fit is statistically preferred
  in respect to the simple power law fit as demonstrated by the F-test
  (probabilities below 5-10\%). More specifically, starting from the
  RLF of FRI galaxies, Urry, Padovani \& Stickel (1991) predict that
  the beamed RLF has negative slopes of around 1.6 for luminosities
  between $10^{21} - 10^{25}$ W/Hz, and 3.3 for the range $4.7\times
  10^{27} - 10^{29}$W/Hz. It is remarkable that without any parameter
  fixing, these values are in such relatively good agreement with
  those found for our fits of the RLF for the Type0/WLAGN (see Table
  \ref{lffit.tab} and Figure \ref{RLFcomp.fig}). Based on their
  simulations, Urry, Padovani \& Stickel (1991) find a space density
  for BL Lacs ($H_{0}=50$~Km~s$^{-1}$Mpc$^{-1}$) is $N \sim 5.1\times
  10^{3}$ Gpc$^{-3}$ for sources with $L > 5 \times10^{20}$W/Hz.

\item The comparison between the RLF derived from the CBS samples and
the prediction from the beaming model are also shown in Figure
\ref{RLFcomp.fig}. Since the high luminosity end of the RLF is
critically dependent on the adopted value for the evolutionary
parameter, we have used the same parameter $C=3$ used by Urry \
Padovani (1995) to derive the beaming parameters. We have also applied
the correction factor of 3/2 in the normalization discussed in
Padovani et al. (2007) to account for the different classification
criteria adopted. What we observe is that the RLF of the Type0 and
WLAGN follow nicely the predicted (but never observed before) shape of
the LF, at least down to $L \sim 10^{23}$W/Hz. A significant departure
from the prediction is only observed in the low end of the luminosity
range ($L\sim 10^{22} - 10^{23}$W/Hz). Such a feature can imply that
either the parent population of BL Lacs was not sampled correctly at
the low end of the luminosity range at the time of Urry \& Padovani's
work, or that the low luminosity end of the CBS RLF is contaminated by
"interlopers" (a situation already discussed in the Caccianiga \&
March\~a, 2004). Deeper observations of these very low-power sources
(only about 5 objects in total) are thus required. In any case, the
presence of the luminosity break (at much higher luminosities) is not
affected by this low $L$ tail.

It is interesting to note that although BL Lac objects with radio
luminosities in the range between $10^{23}-10^{24}$W/Hz are predicted
by the beaming model, no 'classical' BL Lacs are actually found, only
Type0 or WLAGN. This, we believe is a convincing indication that the
usual (classical) BL Lac definition, simply does not work in this
range of lower radio luminosities, and that PEGs or WLAGN should be
considered as BL Lacs.

\item We have investigated how the assumption we made for the z-less
sources affects the predicted RLF by fitting the LF each of the three
random $z$ distributions resulting from the three different L intervals
mentioned in Section 3. The results showed that the effect of changing
the distribution of the z-less sources was to shift the value of
$L_{break}$. For example, for the case of the Type~0 sources we found
that the $Log L_{break}$ varied between 25.01 and 25.67 for the case
of evolution and $H_{0}=50$~Km~s$^{-1}$Mpc$^{-1}$. For the same
scenario but for the WLAGN sample we found that the variation of $Log
L_{break}$ due to the range in the $z$-less distribution (e.g the
luminosity intervals used to derive the redshift distribution for the
$z$-less sources L0-L2 as defined in Section 3.) was between 24.64 and
24.99.

\end{enumerate}

\section{Summary and Conclusions}

We have used a well defined sample of low power flat radio spectrum
sources from the CLASS catalogue with the prime objective of studying
the radio luminosity function of BL Lacs. This objective lead us to
investigate not only the 'classical' BL Lacs found in the CBS, but
also their closest relatives, ie, the other flat radio spectrum
sources where optical line emission is clearly weaker than that found
in quasars. The present investigation builds on knowledge gained from
the extensive work carried out on a pilot sample, the so called
'200mJy sample' (March\~a et al., 1996) - see for instance Bondi et al., 2004,
Dennett-Thorpe \& March\~a, 2000, Ant\'on et al., 2004 - and which
supports the idea that at least some of the weak lined galaxies with
radio properties similar to the classical BL Lacs, are indeed weak BL
Lac nuclei which reside in the nucleus of bright ellipticals.

Hence, the aim of the current study was twofold: to scrutinise the RLF
of BL Lacs at low radio powers whilst investigating the sources that
lie near the cut-off limits of 'classical' BL Lac definition. Allowing
for more physically meaningful selection criteria, and in accordance
with previous works, we analysed three subsamples from the CBS: the
'classical' BL Lacs, the Type0 and the WLAGN.

The conclusions can be summarised as below:

\begin{enumerate}

\item We present one of the deepest and largest samples of BL Lacs
selected at radio frequencies. This gives the unprecedented
opportunity to explore the BL Lac population to radio luminosities
over 100 times lower than existing samples. Furthermore, the work
also greatly improves the statistics for the samples of these types of
sources (nearly a factor of 4).

\item Due to the improved statistics we are able to detect a weak
positive evolution (only marginally detected in previous studies) with
a 2-3$\sigma$ confidence.

\item Reaching lower radio luminosities has enabled the detection of
the change in the slope of the RLF predicted by the unified schemes at
roughly $10^{25}$W/Hz, but never witnessed before. The agreement between
the observed RLF for the Type0 and WLAGN sources of the CBS and the model
predictions are remarkable considering that there was no fitting
involved.

\end{enumerate} 

\section*{Acknowledgments}

The authors would like to thank I. Browne and R. Della Ceca for
helpful comments and suggestions. This research has made use of NASA's
Astrophysics Data System Bibliographic Services.

\appendix
\section{Table A1}

\begin{table*}
\caption{Table containing the sources used in this work. Columns are
as follows: (1) Name; (2) Position (J2000); (3) redshift; (4) Radio flux at 5~GHz (mJy); (5)
R magnitude; (6) radio spectral index; (7) Classification: bl for 
'classical' BL Lac, t0 for Type~0 and w for WLAGN.}
\label{table}
\begin{tabular}{c c r r r r r}
\hline\hline
name & Position & z & S$_{5 GHz}$ & R mag & $\alpha_{1.4}^{4.8}$ & classification \\
     & (J2000)  &   &   (mJy)     &       &                       &                \\
(1) & (2) & (3) & (4) & (5) & (6) & (7) \\
\hline
GB6J013631+390623 & 01 36 32.43 +39 05 59.4 & -- &    49 & 16.4 &  0.17 & bl t0 w \\
GB6J021625+400101 & 02 16 24.55 +40 00 35.9 & 0.050 &    49 & 14.4 & -0.12 &   t0 w \\
GB6J025359+362539 & 02 54 00.03 +36 25 51.5 & 0.048 &    63 & 12.4 &  0.22 &   t0 w \\
GB6J065648+560258 & 06 56 47.58 +56 03 09.2 & 0.056 &    67 & 16.0 & -0.14 &     w \\
GB6J070648+592327 & 07 06 49.54 +59 23 14.7 & 0.094 &    49 & 16.3 &  0.04 &     w \\
GB6J070932+501056 & 07 09 34.27 +50 10 56.4 & 0.019 &   126 & 12.3 & -0.03 &   t0 w \\
GB6J071510+452554 & 07 15 10.01 +45 25 55.5 & 0.052 &    98 & 14.5 & -0.21 &   t0 w \\
GB6J072028+370647 & 07 20 28.77 +37 06 45.2 & 0.120 &    36 & 16.9 & -0.86 &   t0 w \\
GB6J072151+712036 & 07 21 53.10 +71 20 36.7 & -- &   859 & 16.0 & -0.14 & bl t0 w \\
GB6J073328+351532 & 07 33 29.58 +35 15 42.4 & 0.177 &    70 & 16.8 &  0.31 & bl t0 w \\
GB6J073758+643048 & 07 37 58.98 +64 30 43.4 & 0.170 &   251 & 17.1 &  0.41 &   t0 w \\
GB6J073933+495449 & 07 39 34.91 +49 54 39.0 & 0.061 &    55 & 16.1 &  0.49 &     w \\
GB6J080624+593059 & 08 06 25.94 +59 31 06.8 & 0.300 &    38 & 17.3 &  0.38 & bl t0 w \\
GB6J080949+521856 & 08 09 49.23 +52 18 58.1 & 0.138 &   184 & 15.7 & -0.01 & bl t0 w \\
GB6J081622+573858 & 08 16 22.68 +57 39 09.2 & -- &    64 & 17.4 &  0.36 & bl t0 w \\
GB6J083055+540041 & 08 31 00.36 +54 00 24.4 & 0.062 &    40 & 15.3 & -0.71 &     w \\
GB6J083140+460800 & 08 31 39.80 +46 08 00.2 & 0.133 &    98 & 16.2 &  0.24 &   t0 w \\
GB6J083411+580318 & 08 34 11.06 +58 03 21.4 & 0.093 &    57 & 15.7 &  0.01 &     w \\
GB6J083901+401608 & 08 39 03.22 +40 15 46.6 & 0.196 &    34 & 16.8 &  0.25 &   t0 w \\
GB6J085005+403610 & 08 50 04.76 +40 36 07.4 & 0.267 &   110 & 17.3 &  0.15 &     w \\
GB6J090536+470603 & 09 05 36.56 +47 05 46.5 & 0.174 &    86 & 17.1 &  0.14 &   t0 w \\
GB6J090615+463633 & 09 06 15.58 +46 36 19.0 & 0.085 &   159 & 16.6 &  0.49 &     w \\
GB6J090650+412426 & 09 06 52.78 +41 24 29.1 & 0.028 &    65 & 14.5 & -0.16 &   t0 w \\
GB6J090757+493558 & 09 07 56.36 +49 35 48.3 & 0.035 &    36 & 15.0 &  0.14 &   t0 w \\
GB6J092914+501323 & 09 29 15.53 +50 13 35.5 & 0.370 &   544 & 17.3 & -0.03 & bl t0 w \\
GB6J093737+723106 & 09 37 31.93 +72 30 55.2 & 0.114 &    39 & 16.4 & -0.34 &   t0 w \\
GB6J094319+361447 & 09 43 19.16 +36 14 52.2 & 0.022 &    81 & 12.9 & -0.06 &     w \\
GB6J094542+575739 & 09 45 42.29 +57 57 46.2 & 0.229 &    92 & 16.9 &  0.25 & bl t0 w \\
GB6J094557+461907 & 09 45 57.17 +46 19 18.6 & 0.096 &    31 & 16.7 & -0.11 &     w \\
GB6J094832+553538 & 09 48 32.03 +55 35 35.6 & 0.118 &    34 & 15.8 & -0.07 &   t0 w \\
GB6J095847+653405 & 09 58 47.22 +65 33 54.3 & 0.368 &  1125 & 16.6 & -0.35 & bl t0 w \\
GB6J100712+502346 & 10 07 10.42 +50 23 55.8 & 0.133 &    34 & 16.5 & -0.08 &   t0 w \\
GB6J101244+423009 & 10 12 44.22 +42 29 56.7 & 0.366 &    46 & 17.4 &  0.44 & bl t0 w \\
GB6J101504+492606 & 10 15 04.13 +49 26 01.1 & 0.200 &   299 & 16.7 &  0.19 & bl t0 w \\
GB6J101859+591126 & 10 18 58.61 +59 11 27.9 & -- &    76 & 17.3 &  0.21 & bl t0 w \\
GB6J103118+505350 & 10 31 18.51 +50 53 36.9 & 0.361 &    34 & 16.8 &  0.09 & bl t0 w \\
GB6J103318+422228 & 10 33 18.23 +42 22 37.2 & 0.211 &    34 & 17.3 &  0.24 & bl t0 w \\
GB6J103653+444832 & 10 36 52.95 +44 48 18.6 & 0.127 &    66 & 17.3 & -0.37 &     w \\
GB6J103742+571158 & 10 37 44.28 +57 11 56.4 & -- &   126 & 17.4 & -0.46 & bl t0 w \\
GB6J103951+405557 & 10 39 51.04 +40 55 41.8 & 0.075 &    48 & 16.2 & -0.42 &   t0 w \\
GB6J104630+544953 & 10 46 28.73 +54 49 45.3 & 0.249 &    77 & 16.9 &  0.05 &     w \\
GB6J105115+464439 & 10 51 15.97 +46 44 17.7 & 0.100 &   270 & 17.5 &  0.14 &   t0 w \\
GB6J105203+424203 & 10 52 04.15 +42 41 53.3 & 0.136 &    70 & 17.0 &  0.26 &     w \\
GB6J105344+493006 & 10 53 44.02 +49 29 55.2 & 0.140 &    59 & 15.7 &  0.07 & bl t0 w \\
GB6J105430+385500 & 10 54 31.80 +38 55 21.6 & 1.363 &    56 & 17.3 &  0.08 & bl t0 w \\
GB6J105730+405631 & 10 57 31.12 +40 56 46.5 & 0.025 &    31 & 11.7 &  0.30 &     w \\
GB6J105837+562817 & 10 58 37.71 +56 28 11.8 & 0.144 &   247 & 16.1 & -0.06 & bl t0 w \\
GB6J110428+381228 & 11 04 27.42 +38 12 32.9 & 0.030 &   723 & 13.3 &  0.10 & bl t0 w \\
GB6J110508+465311 & 11 05 07.05 +46 53 18.9 & 0.112 &    44 & 17.1 &  0.25 &     w \\
GB6J110552+394649 & 11 05 53.69 +39 46 55.1 & 0.099 &    53 & 17.0 & -0.13 &     w \\
GB6J110657+603345 & 11 06 56.76 +60 34 02.4 & 0.128 &    31 & 16.5 &  0.04 &   t0 w \\
GB6J110939+383046 & 11 09 39.19 +38 31 21.5 & 0.119 &    58 & 17.0 & -0.93 & bl t0 w \\
GB6J111206+352707 & 11 12 08.06 +35 27 06.8 & 0.025 &    49 & 13.0 &  0.22 &   t0 w \\
GB6J111912+623938 & 11 19 16.56 +62 39 26.6 & 0.110 &    43 & 15.5 &  0.06 &   t0 w \\
GB6J112047+421206 & 11 20 48.06 +42 12 14.3 & 0.124 &    30 & 17.3 & -0.18 & bl t0 w \\
GB6J112157+431459 & 11 21 56.66 +43 14 58.8 & 0.185 &    37 & 17.3 &  0.22 &     w \\
GB6J112413+513350 & 11 24 13.14 +51 33 50.1 & 0.234 &    39 & 17.0 &  0.28 & bl t0 w \\
GB6J113626+700931 & 11 36 26.48 +70 09 25.8 & 0.046 &   267 & 13.9 &  0.20 & bl t0 w \\
GB6J113629+673707 & 11 36 29.92 +67 37 06.0 & 0.135 &    47 & 16.6 & -0.02 & bl t0 w \\
GB6J114115+595309 & 11 41 16.00 +59 53 09.2 & 0.012 &   147 & 14.2 & -0.17 &   t0 w \\
\hline
\end{tabular}
\end{table*}
\newpage
\addtocounter{table}{-1}
\begin{table*}
\caption{continue}
\begin{tabular}{c c r r r r r}
\hline\hline
name & Position & z & S$_{5 GHz}$ & R mag & $\alpha_{1.4}^{4.8}$ & classification \\
     & (J2000)  &   &   (mJy)     &       &                       &                \\
\hline\hline
GB6J114300+730413 & 11 43 04.73 +73 04 09.3 & 0.123 &    32 & 16.9 &  0.34 &   t0 w \\
GB6J114312+612214 & 11 43 12.10 +61 22 11.1 & -- &    94 & 17.5 & -0.09 & bl t0 w \\
GB6J114850+592459 & 11 48 50.42 +59 24 57.3 & 0.011 &   566 & 14.2 & -0.13 &     w \\
GB6J114959+552832 & 11 50 00.15 +55 28 21.8 & 0.139 &    83 & 16.8 &  0.50 &   t0 w \\
GB6J115126+585913 & 11 51 24.66 +58 59 18.6 & -- &   131 & 17.4 &  0.28 & bl t0 w \\
GB6J120209+444452 & 12 02 08.43 +44 44 20.8 & 0.298 &    69 & 17.5 &  0.35 & bl t0 w \\
GB6J120304+603130 & 12 03 03.55 +60 31 19.4 & 0.066 &   182 & 16.0 &  0.04 & bl t0 w \\
GB6J120922+411938 & 12 09 22.82 +41 19 41.1 & -- &   459 & 17.1 & -0.42 & bl t0 w \\
GB6J121008+355224 & 12 10 08.05 +35 52 42.4 & 0.022 &    44 & 15.4 & -0.44 &     w \\
GB6J121331+504446 & 12 13 29.28 +50 44 30.3 & 0.031 &    86 & 12.5 &  0.10 &     w \\
GB6J122208+581427 & 12 22 09.40 +58 14 21.5 & 0.100 &    51 & 16.9 &  0.02 &     w \\
GB6J123012+470031 & 12 30 11.80 +47 00 23.0 & 0.039 &    73 & 13.0 &  0.21 &     w \\
GB6J123132+641421 & 12 31 31.34 +64 14 19.4 & 0.170 &    36 & 16.5 &  0.40 & bl t0 w \\
GB6J123151+353929 & 12 31 51.76 +35 39 59.3 & 0.136 &    32 & 16.6 &  0.16 &   t0 w \\
GB6J124313+362755 & 12 43 12.70 +36 27 45.1 & -- &    91 & 17.4 &  0.39 & bl t0 w \\
GB6J124732+672322 & 12 47 33.31 +67 23 16.8 & 0.107 &   174 & 16.6 &  0.34 &   t0 w \\
GB6J124818+582029 & 12 48 18.77 +58 20 28.8 & 0.847 &   356 & 15.8 & -0.30 & bl t0 w \\
GB6J125311+530113 & 12 53 11.94 +53 01 12.1 & -- &   363 & 17.1 &  0.24 & bl t0 w \\
GB6J130132+463357 & 13 01 32.61 +46 34 03.4 & 0.206 &   155 & 16.7 & -0.47 &     w \\
GB6J130146+441612 & 13 01 46.35 +44 16 19.9 & -- &    47 & 17.4 &  0.18 & bl t0 w \\
GB6J130836+434405 & 13 08 37.90 +43 44 15.8 & 0.035 &    47 & 12.6 &  0.18 &     w \\
GB6J130924+430502 & 13 09 25.58 +43 05 05.5 & -- &    45 & 17.0 &  0.23 & bl t0 w \\
GB6J131739+411538 & 13 17 39.18 +41 15 46.4 & 0.067 &   195 & 13.4 &  0.25 &   t0 w \\
GB6J132513+395610 & 13 25 13.34 +39 55 53.7 & 0.075 &    53 & 14.5 &  0.07 &     w \\
GB6J134139+371653 & 13 41 38.81 +37 16 45.3 & 0.170 &    77 & 17.1 &  0.38 &   t0 w \\
GB6J134856+395904 & 13 48 55.95 +39 59 07.3 & 0.008 &    62 & 12.1 &  0.24 &   t0 w \\
GB6J135313+350912 & 13 53 14.28 +35 08 48.3 & 0.139 &    31 & 16.3 &  0.27 &   t0 w \\
GB6J135327+401700 & 13 53 26.67 +40 16 58.8 & 0.008 &    35 & 10.9 &  0.13 &     w \\
GB6J141132+742404 & 14 11 34.73 +74 24 29.4 & -- &    82 & 17.4 &  0.21 & bl t0 w \\
GB6J141343+433959 & 14 13 43.68 +43 39 45.5 & 0.089 &    39 & 16.1 &  0.18 &     w \\
GB6J141536+483102 & 14 15 36.77 +48 30 30.6 & 0.496 &    37 & 17.4 &  0.10 & bl t0 w \\
GB6J141946+542328 & 14 19 46.50 +54 23 15.1 & 0.151 &  1350 & 15.5 & -0.41 & bl t0 w \\
GB6J151717+694715 & 15 17 14.61 +69 47 10.2 & 0.137 &    32 & 17.5 & -0.64 &     w \\
GB6J151746+652456 & 15 17 47.55 +65 25 23.6 & 0.702 &    31 & 17.4 &  0.16 & bl t0 w \\
GB6J151807+665746 & 15 18 08.95 +66 57 53.4 & 0.057 &    36 & 12.9 &  0.06 &     w \\
GB6J151838+404532 & 15 18 38.93 +40 45 00.7 & 0.065 &    44 & 15.4 &  0.01 &     w \\
GB6J153900+353053 & 15 39 01.66 +35 30 46.1 & 0.080 &    92 & 16.5 &  0.04 &     w \\
GB6J154255+612950 & 15 42 56.94 +61 29 55.5 & 0.507 &   121 & 17.3 & -0.27 & bl t0 w \\
GB6J155848+562524 & 15 58 48.30 +56 25 14.4 & 0.300 &   206 & 17.3 &  0.01 &   t0 w \\
GB6J155901+592437 & 15 59 01.67 +59 24 21.5 & 0.060 &   191 & 12.9 &  0.11 &   t0 w \\
GB6J162509+405345 & 16 25 10.35 +40 53 34.4 & 0.030 &   110 & 12.2 &  0.32 &   t0 w \\
GB6J164420+454644 & 16 44 20.05 +45 46 45.4 & 0.223 &   109 & 17.1 &  0.44 & bl t0 w \\
GB6J164734+494954 & 16 47 35.13 +49 49 57.2 & 0.048 &   191 & 16.5 & -0.04 &     w \\
GB6J165353+394541 & 16 53 52.24 +39 45 36.6 & 0.034 &  1375 & 11.5 &  0.10 & bl t0 w \\
GB6J165547+444735 & 16 55 47.36 +44 47 25.0 & 0.076 &    35 & 15.2 &  0.07 &     w \\
GB6J170123+395432 & 17 01 24.70 +39 54 36.2 & -- &   150 & 17.3 &  0.19 & bl t0 w \\
GB6J170449+713840 & 17 04 47.05 +71 38 16.9 & 0.350 &    43 & 16.8 & -0.15 & bl t0 w \\
GB6J171523+572434 & 17 15 22.94 +57 24 40.5 & 0.027 &    35 & 11.3 &  0.40 &   t0 w \\
GB6J171718+422711 & 17 17 19.18 +42 26 59.6 & 0.183 &   125 & 16.0 &  0.06 &     w \\
GB6J172535+585127 & 17 25 35.07 +58 51 39.3 & 0.297 &    55 & 17.1 &  0.26 & bl t0 w \\
GB6J172722+551059 & 17 27 23.49 +55 10 53.9 & 0.247 &   274 & 17.3 & -0.52 &     w \\
GB6J172818+501315 & 17 28 18.58 +50 13 11.3 & 0.055 &   145 & 15.7 &  0.37 & bl t0 w \\
GB6J173047+371451 & 17 30 46.88 +37 14 55.0 & -- &    78 & 17.2 &  0.22 & bl t0 w \\
GB6J173410+421933 & 17 34 13.53 +42 19 57.3 & 0.267 &    36 & 15.3 &  0.28 & bl t0 w \\
GB6J174113+722447 & 17 41 22.62 +72 24 51.5 & 0.220 &    52 & 17.5 &  0.43 & bl t0 w \\
GB6J174231+594513 & 17 42 31.96 +59 45 07.3 & 0.400 &    88 & 17.1 &  0.16 & bl t0 w \\
GB6J174832+700550 & 17 48 32.88 +70 05 51.6 & 0.770 &   715 & 16.9 &  0.02 & bl t0 w \\
GB6J174900+432151 & 17 49 00.21 +43 21 51.8 & -- &   321 & 17.5 & -0.11 & bl t0 w \\
GB6J175041+395706 & 17 50 41.17 +39 57 00.3 & 0.049 &    38 & 14.5 &  0.08 &   t0 w \\
GB6J175546+623652 & 17 55 48.36 +62 36 44.4 & 0.027 &   203 & 11.0 &  0.28 &     w \\
\hline
\end{tabular}
\end{table*}
\newpage
\addtocounter{table}{-1}
\begin{table*}
\caption{continue}
\begin{tabular}{c c r r r r r}
\hline\hline
name & Position & z & S$_{5 GHz}$ & R mag & $\alpha_{1.4}^{4.8}$ & classification \\
     & (J2000)  &   &   (mJy)     &       &                       &                \\
\hline\hline
GB6J175628+580708 & 17 56 29.19 +58 06 58.2 & 0.192 &    38 & 16.9 &  0.25 &     w \\
GB6J175704+535153 & 17 57 06.74 +53 51 37.3 & 0.119 &    42 & 15.1 &  0.29 &   t0 w \\
GB6J175728+552309 & 17 57 28.37 +55 23 11.7 & 0.065 &    73 & 14.1 &  0.07 &   t0 w \\
GB6J180651+694931 & 18 06 50.46 +69 49 28.1 & 0.051 &  2122 & 15.1 & -0.04 & bl t0 w \\
GB6J180738+563159 & 18 07 37.75 +56 31 56.8 & 0.059 &    32 & 15.5 & -0.21 &   t0 w \\
GB6J183850+480237 & 18 38 49.22 +48 02 35.0 & 0.300 &    41 & 17.4 & -0.25 & bl t0 w \\
GB6J183858+573535 & 18 38 58.66 +57 35 38.1 & 0.164 &    93 & 16.9 & -0.05 & bl t0 w \\
GB6J184033+621257 & 18 40 33.53 +62 12 49.5 & 0.050 &    72 & 16.5 & -0.06 &   t0   \\
GB6J191212+660826 & 19 12 07.38 +66 07 46.9 & 0.075 &    39 & 15.1 &  0.44 &     w \\
GB6J230115+351252 & 23 01 14.45 +35 13 00.6 & 0.136 &   129 & 16.8 & -0.29 &   t0 w \\
\hline
\end{tabular}
\end{table*}


\vspace*{1cm}


\begin{thebibliography}{99}

\bibitem{b1} Ant\'on, S. \& Browne, I.W.A., 2005, MNRAS, 356, 225.
\bibitem{b1} Ant\'on, S., Browne, I.W.A., March\~a, M.J.M., 2004,
MNRAS, 352, 673.
\bibitem{b1} Avni Y., Bahcall J.N., 1980, ApJ, 235, 694.
\bibitem{b1} Bade N., Beckmann V., Douglas N.G., Barthel P.D., Engels D.,
Cordis L., Nass P., Voges W., 1998, A\&A, 334, 459.

\bibitem{b1} Beckmann, V., Engels, D., Bade, N., Wucknitz, O., 2003, A\&A, 401, 927.


\bibitem{b1} Bondi, M., March\~a, M.J.M., Polatidis, A.,
Dallacasa, D., Stanghellini, C., Ant\'on, S., 2004, MNRAS, 352, 112.
\bibitem{b1} Browne I.W.A., March\~a M.J.M. 1993, MNRAS, 261, 795
\bibitem{b1} Caccianiga A., Maccacaro T., Wolter A., Della Ceca R.,
Gioia, I.M. 1999, ApJ, 513, 51

\bibitem{b1} Caccianiga A., Maccacaro T., Wolter A., Della Ceca R.,
Gioia, I.M. 2002b, ApJ, 566, 181.

\bibitem{b1} Caccianiga A., March\~a M.J.M., Ant\'on A., Mack K.-H., Neeser M.,
2002a, MNRAS, 329, 877. (paper~II)


\bibitem{b1} Caccianiga A., March\~a M.J.M., 2004, MNRAS, 348, 937.

\bibitem{b1} Capetti, A., Celotti, A., Chiaberge, M., de Ruiter,
H. R., Fanti, R., Morganti, R., Parma, P., 2002, A\&A, 383, 104.

\bibitem{b1} Dennett-Thorpe, J.,  March\~a, M. J.,  2000, A\&A, 361, 480.

\bibitem{b1} Giommi P., Menna M.T., Padovani P., 1999, MNRAS, 310, 465
\bibitem{b1} Giommi P., Pellizzoni A., Perri M., Padovani P., 2001, in proc.
``Blazar Dempgraphics and Physics'', eds. P. Padovani \& C.M. Urry, 
ASP Conf. Ser., vol. 227, p.227

\bibitem{b1} Giommi P., Padovani P., Polenta, G., Turriziani, S.,
  D'Elia, V., Piranomonte, S., 2012, MNRAS, 420, 2899.

\bibitem{b1} Kaiser, C.R. and Best, P. N., 2007, MNRAS, 381, 1548.

\bibitem{b1} Landt, H., Padovani, P., Giommi, P., 2002, MNRAS, 336, 445.

\bibitem{b1} Landt, H., Padovani, P., Perlman, E. S., Giommi, P.,
  2004, MNRAS, 351, 83.

\bibitem{b1} Laurent-Muehleisen S.A., Kollgaard R.I., Ciardullo R., 
Feigelson E.D., Brinkman W., Siebert J., 1998, ApJS, 118, 127

\bibitem{b1} Laurent-Muehleisen S.A., Kollgaard R.I., Feigelson E.D.,
  Brinkman W., Siebert J., 1999, ApJ, 525, 127


\bibitem{b1} March\~a M.J.M., Browne I.W.A., Impey C.D., Smith P.S., 1996,
MNRAS, 281, 425
\bibitem{b1} March\~a M.J.M., Browne I.W.A. 1995, MNRAS, 275, 951
\bibitem{b1} March\~a M.J.M., Caccianiga A., Browne I.W.A., Jackson N., 
2001, MNRAS, 326, 1455 (paper~I).

\bibitem{b1} Massaro, E., Giommi, P., Leto, C., Marchegiani, P.,
Maselli, A., Perri, M., Piranomonte, S., 2011, 'Multifrequency
Catalogue of Blazars (3rd Edition)', eds. Massaro, E., E., Giommi, P.,
Leto, C., Marchegiani, P., Maselli, A., Perri, M., Piranomonte, S.,
ARACNE Editrice, Rome, Italy.

\bibitem{b1} Meisner, A. M., Romani, R. W., 2010, ApJ, 712, 14.

\bibitem{b1} Morris S.L., Stocke J.T., Gioia I.M., Schild R.E., 
Wolter A., Maccacaro T., della Ceca R., 1991, ApJ, 380, 49

\bibitem{b1} Padovani P., Giommi P., Landt, H., Perlman, E. S., 2007, ApJ, 662, 182. 

\bibitem{b1} Perlman E.S., Stocke J.T., Schachter J.F., Elvis M.,
 Ellingson E., Urry C.M., Potter M., Impey C.D., Kolchinsky, P. 1996, 
ApJS, 104, 251 

\bibitem{b1} Perlman E.S., Padovani, P., Giommi, P., Sambruna, R.,
  Jones, L.R., Tzioumis, A., \& Reynolds, J., 1998, AJ, 115, 1253.


\bibitem{b1} Rau, A., Schady, P., Greiner, J. Salvato, M., Ajello, M.,
  Bottacini, E., Geherls, N., Afonso, P. M. J., Elliott, J., Filgas,
  R., (and 10 other authors), 2012, A\&A, 538, 26.
\bibitem{b1} Rector T.A., Stocke J.T., Perlman E.S., Morris S.L., Gioia I.M.,
2000, AJ, 120, 1626 

\bibitem{b1} Sbarufatti, B., Cipriani, S., Kotilainen, J., Deacrli,
R., Treves, A., Veronesi, A. Falomo, R., 2009, AJ, 137, 337.

\bibitem{b1} Schmidt M., 1968, ApJ, 51, 393
\bibitem{b1} Stickel M., Fried J.W., K\"uhr H., Padovani P., Urry C.M.,
1991, ApJ, 374, 431
\bibitem{b1} Stocke J.T., Morris S.L., Gioia I.M., Maccacaro T., 
Schild R., Wolter A., Fleming T.A., Henry J.P. 1991, ApJS, 76, 813

\bibitem{b1} Trussoni, E., Capetti, A., Celotti, A., Chiaberge, M.,
Feretti, L., 2003, A\&A, 403, 889.

\bibitem{b1} Urry C.M., Padovani P., 1991, ApJ, 371, 60

\bibitem{b1} Urry, C. M., Padovani, P., Stickel, M., 1991, ApJ, 382, 501.

\bibitem{b1} Urry C.M., Padovani P., 1995, PASP, 107, 803
\bibitem{b1} Urry C.M., Shafer R.A., 1984, ApJ, 280, 569

\bibitem{b1} Wolfe, M., 1978. Pittsburgh Conference on BL Lacs (ed.)
  Arthur M Wolfe, Published Universtity of Pittsburgh, Department of
  Physics and Astronomy, Pittsbught, PA, 15620.

\bibitem{b1} Wolter, A., Caccianiga, A., Della Ceca, R., and
Maccacaro, T., 1994, ApJ, 433, 29.

\end{thebibliography}
\end{document}